# Are mouse and cat the missing link in the COVID-19 outbreaks in seafood markets?

Daniel H. Tao[1], Weitao Sun[2,3*]

[1] The High School Affiliated to Remin University of China, Beijing, 100080, China

[2] School of Aerospace Engineering, Tsinghua University, Beijing, 100084, China

[3] Zhou Pei-Yuan Center for Applied Mathematics, Tsinghua University, Beijing, 100084, China

*Corresponding author:   Weitao Sun

Email: sunwt@tsinghua.edu.cn

## Abstract

Severe acute respiratory syndrome coronavirus 2 (SARS-CoV-2) virus caused the novel coronavirus disease-2019 (COVID-19) affecting the whole world. Like SARS-CoV and MERS-CoV, SARS-CoV-2 are thought to originate in bats and then spread to humans through intermediate hosts. Identifying intermediate host species is critical to understanding the evolution and transmission mechanisms of COVID-19. However, determining which animals are intermediate hosts remains a key challenge. Virus host-genome similarity (HGS) is an important factor that reflects the adaptability of virus to host. SARS-CoV-2 may retain beneficial mutations to increase HGS and evade the host immune system. This study investigated the HGSs between 399 SARS-CoV-2 strains and 10 hosts of different species, including bat, mouse, cat, swine, snake, dog, pangolin, chicken, human and monkey. The results showed that the HGS between SARS-CoV-2 and bat was the highest, followed by mouse and cat.

Human and monkey had the lowest HGS values. In terms of genetic similarity, mouse and monkey are halfway between bat and human. Moreover, given that COVID-19 outbreaks tend to be associated with live poultry and seafood markets, mouse and cat are more likely sources of infection in these places. However, more experimental data are needed to confirm whether mouse and cat are true intermediate hosts. These findings suggest that animals closely related to human life, especially those with high HGS, need to be closely monitored.

## Introduction

In December 2019, a novel coronavirus was announced and later named as the Severe Acute Respiration Syndrome Coronavirus (SARS-CoV-2). The SARS-CoV-2 infected human through respiration system, causing serious inflammatory reaction and damages of lung[1, 2], kidney[3], digestive system[4] and heart[5], and cause liver damage[6]. Compared with SARS-CoV and MERS-CoV, SARS-CoV-2 evolved an enhanced ability to survive and spread in environments such as food surfaces and packaging. The infection rate of the SARS-CoV-2 is at a very high level, causing an ongoing worldwide pandemic. By August 2020, the total number of confirmed cases globally was about 20,426,316 and the number of deaths caused by SARS-CoV-2 was about 740,143.

It is estimated that about 60% of known infectious diseases and 75% new diseases originate in animals[7-9]. One of the crucial ways to stop the spread of the virus is to find the intermediate hosts of SARS-CoV-2. Bats are believed to be the origin host of several kinds of deadly coronaviruses[10, 11]. Recent studies reported that the SARS-CoV-2 found in human has nearly 96% genome similarity with that in bats[10]. Nevertheless, how the virus jumped from bats to human is still unclear. Studies suggested that pangolins and snakes may be potential intermediate hosts for SARS-CoV-2[12-15]. But further investigation is required to make a final conclusion[16, 17].

It has been reported that live poultry markets may be the source of SARS-CoV and influenza[18]. There are hundreds of stalls selling all kinds of meat, seafood, live poultry, and sometimes wild game. Live poultry markets are thought to play an important role in influenza epidemiology[19]. Human influenza virus (H5N1) had clear genetic similarity with avian influenza that preceded human infection[20]. And most human patients infected by avian influenza A (H7N9) virus were directly exposed to infected poultry or indirect exposure in contaminated environments[21]. There is widespread concern that live poultry markets present a risk of cross-infection between humans and animals.

The early discovery of SARS-CoV-2 in December 2019 was related to a seafood market. On 10 June, the new outbreak of COVID-19 in Beijing was also linked to a large market, particularly stalls in the seafood market. The SARS-CoV-2 has been found on salmon chopping boards, and even at the surface of soybean products. Recent reports indicate that SARS-CoV-2 was also found on the packaging surfaces of frozen seafood[22, 23]. The chances of getting infected through exposure to live poultry, seafood and other products are real. But the link between SARS-CoV-2 and seafood markets is still a mystery.

It is hypothesis that people working at the seafoods market may be a potential source of SARS-CoV-2[22], which implies that there were unidentified human host triggered the first outbreak of COVID-19. Repeated outbreaks in live poultry and seafood markets suggest a specific pattern of virus transmission that we do not yet understand. In open seafood markets with many stalls, animals frequently come into contact with humans and food, which makes it very likely that animals can carry and spread the virus. It is not unreasonable to suggest that SARS-CoV-2 may be related to animal pollutions in the market's environment before the virus can spread across species on a large scale.

The coronaviruses jump from bats to human through intermediate hosts, such as palm civets and dromedary camels. A recent study on host range showed that many different species, including mouse, swine, chicken, domestic cat and tiger, pangolin, mink and snake, could potentially be infected by the SARS-CoV-2[24-26]. There is high chance that occurrence of SARS-CoV-2 in seafood markets might be related to common animals, such as cats, mice or dogs. But the roles such mammals played in the spreading process of the SARS-CoV-2 have not been fully understood.

In this study, we proposed a method to search for possible intermediate hosts of SARS-CoV-2 based on the determination of host-genome similarity (HGS)[27]. The exchange of genetic information between virus and hosts can help to enhance the viral adaption, which leads to higher host and virus genome similarity. The high genome similarity between virus and host implies that the virus may have stronger infectivity and adaptability to the host[24, 27]. Studying viral HGS with bats, humans and a range of intermediate hosts may shed lights on how SARS-CoV-2 crosses the species barrier in terms of genetic similarity.

We examined the genome similarities of a set of SARS-CoV-2 strains to 10 hosts of different species, including bat, mouse, cat, swine, snake, dog, pangolin, chicken, human and monkey. The virus genome data set consists of 399 coding sequences of SARS-CoV-2 with geolocation from China. The results showed that SARS-CoV-2 had a high degree of similarity to the bat genome, followed by mouse and cat, and then other animals. Human and monkey were at the bottom of the HGS ranking. Based on genomic similarities, mouse and cat were likely the missing link in the viral transmission chain from bat to human. In addition to genetic similarity data, mouse and cat were also most likely to be present at live poultry markets and exposed to seafood and other products. This raises the possibility of mouse and cat as the source of infctions in seafood markets.

Although direct contact between the mouse/cat and affected seafood has not been formally confirmed, the results suggest that the virus may be transmitted to humans from the original host (bat) through intermediate hosts (mouse and cat) with high HGS. This conclusion implies that there may be an overlooked animal transmission chain composed of animals in close contact with human daily life. Humans may be at risk of animal-to-human or animal-to-food-to-human transmissions of SARS-CoV-2. This work shows the importance of enhanced animal surveillance, especially with animals such as mouse and cat that have close contact with humans.

## Results

**SARS-CoV-2 has high HGSs with bat, mouse and cat**

Quantitative measurement of host-genome similarity (HGS) has been used to detect SARS-CoV-2 gene mutations[27]. Here we investigated the HGS between viral coding sequences (CDSs) of SARS-CoV-2 and genomes of 10 hosts, including human (*Homo sapiens* GRCh38.p12 chromosomes), bat (*Rhinolophus ferrumequinum* mRhiFer1_v1.p), mouse (*Mus musculus* GRCm38.p6), cat (*Felis catus* Felis_catus_9.0), swine (*Sus scrofa* Sscrofa11.1), snake (*Pseudonaja textilis* EBS10Xv2-PRI), dog (*Canis lupus* CanFam3.1), pangolin (*Manis javan*ica ManJav1.0), chicken (*Gallus gallus* GRCg6a) and monkey (*Macaca mulatta* Mmul_10). The NCBI Blastn was used to obtain the CDS alignment scores [28]. The HGS between SARS-CoV-2 and hosts was calculated based on the method in a recent work[27].

The SARS-CoV-2 has a positive-sense single-stranded RNA approximately 30 kb in length. Viral RNA sequence contains 14 5'-ORFs (open reading frames) with length no less than 75 nt (Figure 1). These ORFs encode a series of viral proteins: the replication proteins open reading frame ORF1a and ORF1ab, the structural proteins S (spike), E (envelope), M (membrane), N (nucleocapsid) and accessory proteins (such as ORF6, ORF7a, ORF7b, ORF8) with unknown homologues. The coronavirus ORFs are usually organized in a pattern '-S-X-E-M-X-X-X-N-' where 'X' represents accessory protein genes. The number of accessory genes 'X' may vary in different viruses (such as SARS-CoV, MERS-CoV and SARS-CoV-2).

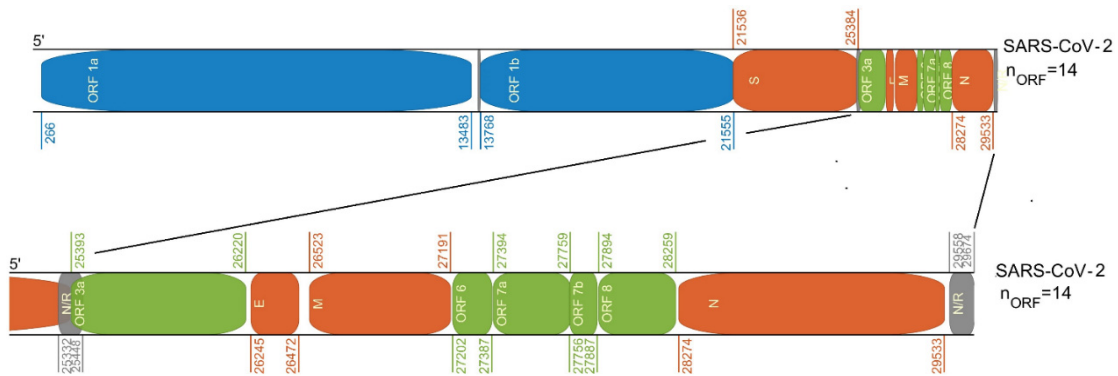

Figure 1. Genome organization of SARS-CoV-2.

By comparing with the host gene sequence, each ORF segment of SARS-CoV-2 has a host-genome similarity. A genome-wide HGS can be obtained by the weighted sum of ORF HGSs. The weight coefficient is the fraction of the ORF length in the entire genome. The HGS values was calculated for 399 SARS-CoV-2 genomes with geolocation from China. Figure 2-4 show the HGS between S (spike) gene, ORF6 and ORF8 of SARS-CoV-2 and the genomes of 10 hosts.

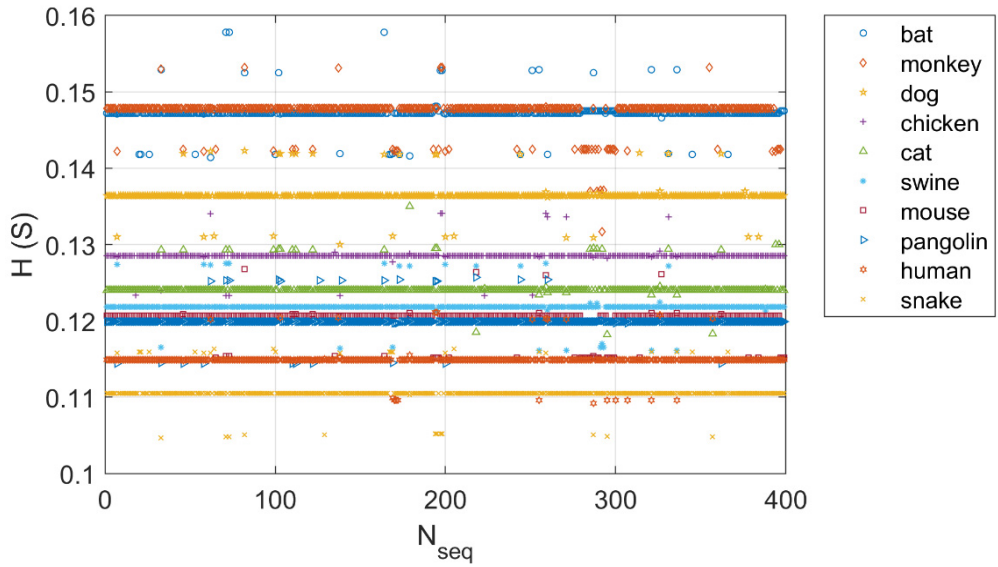

Figure 2. HGS between S (spike) gene and host genomes for 399 SARS-CoV-2 genomes with geolocation from China.

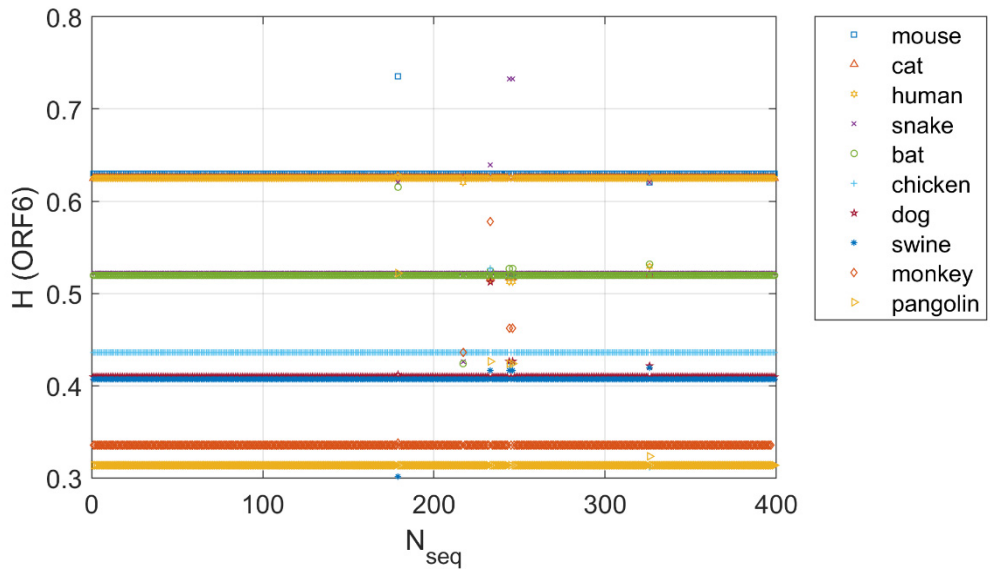

Figure 3. HGS between ORF6 gene and host genomes for 399 SARS-CoV-2 genomes with geolocation from China.

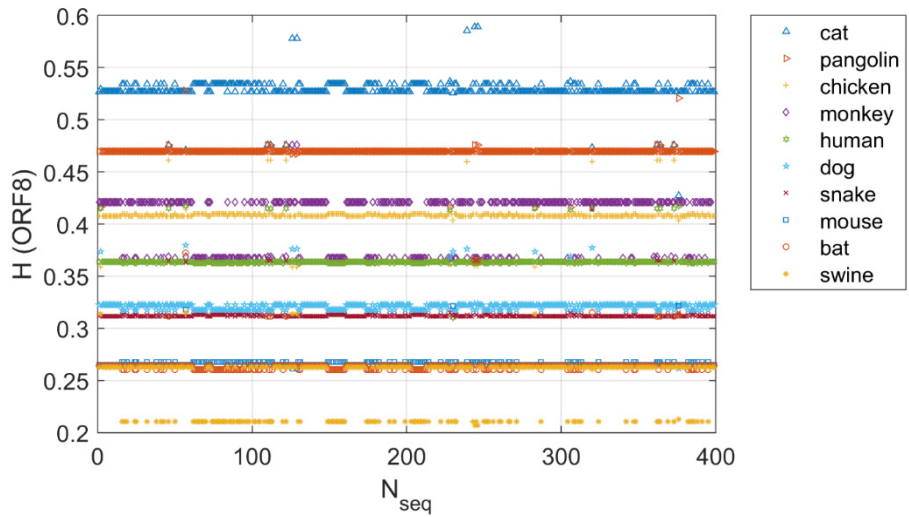

Figure 4. HGS between ORF8 gene and host genomes for 399 SARS-CoV-2 genomes with geolocation from China.

Amazingly, the host-genome similarity of SARS-CoV-2 genes were quite different for different hosts. For the S (spike) gene of SARS-CoV-2, the largest HGS occured between virus and bat (Figure 2). While for ORF6 and ORF8, SARS-CoV-2 had largest HGSs for mouse and cat respectively (Figure 3, Figure 4). The HGS values of the 399 viral genomes were quite consistent for each kind of host. At the same time, the S (spike) gene had much more data fluctuations than that of ORF6 and ORF8. Such data dispersion clearly demonstrated that the S gene contains more genetic variations.

For different viral genes, the ranking of genome similarity to the host was also changing. Since RNA viruses are susceptible to gene mutations, SARS-CoV-2 may evolve improved host adaptability by mutating different ORF sequences. By acquiring immune modulation genes from cells, viral proteins can regulate or inhibit the host's immune system[29, 30]. Under the selection pressure exerted by the host immune system, SARS-CoV-2 may possess the ability to mutate specific genes against different hosts and escape the hosts' innate immune system.

Different genes encode different viral proteins that play various roles during invading host cells. SARS-CoV-2 spike protein binds to angiotensin-converting enzyme 2 (ACE2) on the surface of host cell and translocate into cell through endocytosis process[31]. ORF6 protein of SARS-CoV enhances suppression of IFNβ expression in host innate immunity[32]. As a membrane protein with 63 amino acids, ORF6 also blocks the IFNAR-STAT signaling pathway by limiting the mobility of the importin subunit KPNB1 and preventing the STAT1 complex from moving into the nucleus for ISRE activation[33]. In addition, ORF6 circumvents IFN production by inhibiting IRF-3 phosphorylation in the (TRAF3)-(TBK1+IKKε)-(IRF3)-(IFNβ) signaling pathway, which is an essential signaling pathway triggered by the viral sensors RIG-1/MDA5 and TLRs[34]. ORF8 encodes a single accessory protein at the early stage of SARS-CoV infection and splits into two fragments, ORF8a and ORF8b, at later stages[35]. Proteins ORF8b and ORF8ab in SARS-CoV inhibited the IFN response during viral infection[36]. The antagonism between the virus and host's immune system may have different biological mechanisms in different hosts. From the perspective of ORF genetic similarity, the change of host rank may be a reflection of the different viral adaptabilities to different hosts.

In addition to the similarity between viral ORF and host genome, genome-wide HGS is an effective factor of viral transmission/pathology capability. The full-length genome data were from the Global Initiative on Sharing All Influenza Data (GISAID) database[37]. The sequence requirements were full-length sequences only, sequences with definite collection dates and locations, and no nucleotide names other than A, G, C and T. The number of genomes that met such requirements was 399 for China by the time of 21 June, 2020.

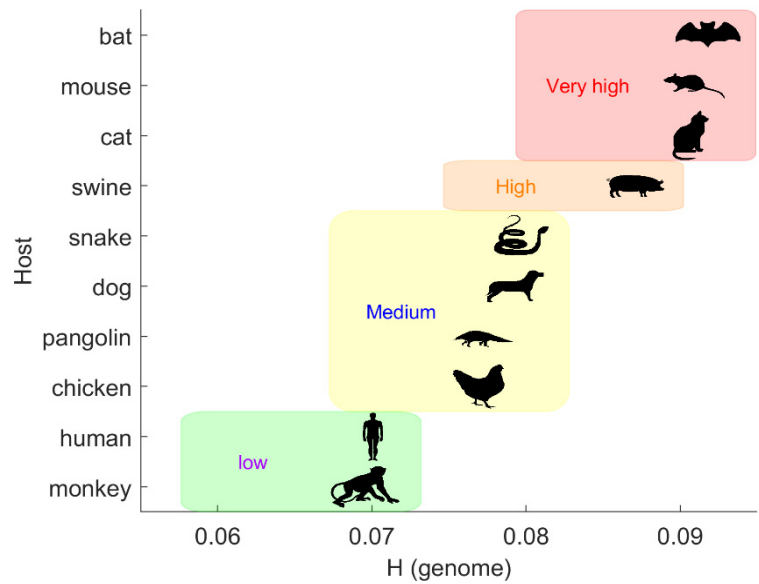

Figure 5. Rank of 10 hosts based on the mean genome-wide HGS of 399 SARS-CoV-2 genomes with geolocation from China. Bats had the highest HGS, while human and monkey had the lowest. In order of HGS values, the intermediate hosts between bat and human are mouse, cat, swine, snake, dog, pangolin and chicken.

Figure 5 shows that SARS-CoV-2 had the highest mean genome-wide HGS with bat, followed by mouse, cat, swine, snake, dog, pangolin, chicken, human and monkey. The mean genome-wide HGS was the average of genome-wide HGS for all 399 SARS-CoV-2 strains (Table 1). Bat, mouse and cat were the top 3 hosts with the highest HGSs (0.09179, 0.09089 and 0.09084 respectively). Human and monkey are hosts that had the lowest HGS (0.07007 and 0.06956 respectively). The mean genome-wide HGS values of swine, snake, dog, pangolin and chicken were significantly lower than those of mice and cats (Figure 5). The roles of mouse and cat in SARS-CoV-2 transmission have not been confirmed. However, by identifying the genome similarity of SARS-CoV-2 with a set of hosts, the mouse and cat had more chances to be potential candidates carrying virus and polluting products in live poultry market inferred from the rank of HGS values.

Table 1. The mean genome-wide HGS between SARS-CoV-2 and 10 hosts

| bat | mouse | cat | swine | snake | dog | pangolin | chicken | human | monkey |
|---|---|---|---|---|---|---|---|---|---|
| 0.09179 | 0.09089 | 0.09084 | 0.08710 | 0.07968 | 0.07920 | 0.07705 | 0.07692 | 0.07007 | 0.06956 |

HGS is a measure of genome similarity between virus and host, which may reflect the adaptability to host's immune system. The identification of intermediate hosts based on mean genome-wide HGS demonstrates that SARS-CoV-2 can become well adapted to different hosts. It is reasonable to hypothesize that animals with high SARS-CoV-2 HGS may be origin or intermediate hosts. The results show that bat had the highest HGS. It has been confirmed that bats were the source of many fatal virus of infectious diseases, including Ebola, rabies, Nipah and Hendra. Mouse and cat also have very high HGS. According to the Center for Disease Control and Prevention, house mice carry at least 7 deadly diseases, such as Hantavirus Pulmonary Syndrome, Leptospirosis, Lymphocytic Choriomeningitis, Plague, Rat-Bite Fever, Tularemia and Salmonellosis. Infected mouse and cat can spread diseases at places from restaurants, grocery stores and live poultry market. The findings suggest that HGS is a way to look for intermediate hosts from a genetic similarity perspective. However, the relationship between HGS and cross-species transmission needs to be carefully interpreted until experimental data are available.

## Methods and materials

### Viral genome data

By using BLAST ORFfinder[38], 14 ORFs were detected in the RNA genome sequence (29903 nt) of SARS-CoV-2 (GenBank: MN908947.3). The lengths of these ORFs were no less than 75 nt and nested

ORFs were ignored. The 399 SARS-CoV-2 genomes (by June 21, 2020) with geolocation as China were from the GISAID database (https://www.gisaid.org/) [37]. Here complete and high-coverage genomes were selected to for accurate HGS determinations. The viral genomes only contain nucleotide A, G, C and T. The MATLAB (https://www.mathworks.com/help/bioinfo/ref/seqshoworfs.html) was used to find SARS-CoV-2 CDSs. The accession IDs of the genomes and ORF HGS values can be found in the Supplemental Information.

**Host-genome similarity (HGS)**

The SARS-CoV-2 CDSs were aligned with the genomes of ten hosts (*Homo sapiens* GRCh38.p12, *Rhinolophus ferrumequinum* mRhiFer1_v1.p, *Mus musculus* GRCm38.p6, *Felis catus* Felis_catus_9.0, *Sus scrofa* Sscrofa11.1, *Pseudonaja textilis* EBS10Xv2-PRI, *Canis lupus* CanFam3.1, *Manis javan*ica ManJav1.0, *Gallus gallus* GRCg6a, *Macaca mulatta* Mmul_10) by Blastn[28]. The HGS is defined as

$$H = \frac{\sum a}{n} = \frac{\sum S'}{2n}, \qquad (1)$$

where $n$ represents the length of the target sequence, $a = S'/2$ and $S' = \frac{\lambda S - \ln K}{\ln 2}$. S is the original score given by Blastn sequence alignment. Parameters K and λ represent the statistical significance of the alignment[39]. Since viral ORFs have quite different lengths, the alignment results between ORF and host genome must be rescaled for comparison. The HGS value $H$ is a measurement of matched base pairs when the genome are rescaled into sequences of the same length.

**Data availability**

The SARS-CoV-2 genomes were obtained at GISAID website (https://www.gisaid.org/). The accession ID and HGS of 399 SARS-CoV-2 genomes are in Supplemental Information. The code for

HGS calculation is available in GitHub (https://github.com/WeitaoNSun/HGS).